\documentclass[a4paper,11pt]{article}
\usepackage{pos}

\title{Phase diagram of the $2+1$-dimensional Gross-Neveu model with chiral imbalance}

\author*[a]{Marc Winstel}
\author[a]{Laurin Pannullo}
\author[a,b]{Marc Wagner}

\affiliation[a]{Institut für Theoretische Physik, Goethe Universit\"at Frankfurt am Main \\
	Max-von-Laue-Str. 1, 60438 Frankfurt am Main, Germany}
\affiliation[b]{Helmholtz Research Academy Hesse for FAIR,\\
  Campus Riedberg, Max-von-Laue-Stra{\ss}e 12, D-60438 Frankfurt am Main, Germany}

\emailAdd{winstel@itp.uni-frankfurt.de}
\emailAdd{pannullo@itp.uni-frankfurt.de}
\emailAdd{mwagner@itp.uni-frankfurt.de}

\abstract{
		In this work, the phase diagram of the $2+1$-dimensional Gross-Neveu model is investigated with baryon chemical potential as well as chiral chemical potential in the mean-field approximation.
		We study the theory using two lattice discretizations, which are both based on naive fermions.
		An inhomogeneous chiral phase is observed only for one of the two discretizations.
		Our results suggest that this phase disappears in the continuum limit.
}

\FullConference{%
	The 38th International Symposium on Lattice Field Theory, LATTICE2021
	26th-30th July, 2021
	Zoom/Gather@Massachusetts Institute of Technology
}
\usepackage{hyperref}
\usepackage[capitalise]{cleveref}
\usepackage{graphicx}
\usepackage{subcaption}
\usepackage{mathtools}
\usepackage{amsmath}
\usepackage{xcolor}
\usepackage{dsfont}
\usepackage{bbold}
\usepackage[utf8]{inputenc}
\usepackage[ngerman, english]{babel}
\usepackage{xstring}

\renewcommand{\ref}[1]{(\ref{#1})}
\newcommand{\xt}[1]{#1}
\newcommand{\x}[1]{\textbf{#1}}
\newcommand{\Ns}{\ensuremath{N_{s}}}
\newcommand{\Nt}{\ensuremath{N_{t}}}
\newcommand{\Nf}{\ensuremath{N_{f}}}
\newcommand{\ii}{\ensuremath{\mathrm{i}}}

\newcommand{\seff}{S_{\text{eff}}}
\newcommand{\kron}[2]{\ensuremath{\delta_{#1,#2}}}

\newcommand{\muff}{\ensuremath{\mu_{45}}}
\newcommand{\diag}{\ensuremath{\mathrm{diag}}}
\providecommand{\Rcite}[1]{%
	\begingroup
	\def\tempx{0}%
	\StrCount{#1}{,}[\tempx]%
	\ifnum\tempx > 0 
	Refs.~%
	\else
	Ref.~%
	\fi
	\endgroup
	\cite{#1}%
}

\definecolor{corr}{rgb}{0,0,1}

\begin{document}
	\maketitle
\section{Introduction}
	The Gross-Neveu (GN) model describes $N_f$ fermion flavors with a quartic interaction in the scalar channel. It is frequently used as a toy model for spontaneous breaking of chiral symmetry \cite{Gross:1974jv}.
	In general, four-Fermi theories are of interest in many applications in high energy physics \cite{Narayanan:2020ahe, Hands:1992ck, Gies:2010st, Scherer:2012nn, Hands:2001cs}, condensed matter physics\footnote{Especially $2+1$-dimensional models are often used to describe superconducting electrons in high-temperature superconductors, which are confined to planes.} \cite{Chodos:1993mf, Zhukovsky:2013vva,Ohsaku:2003rq, MacKenzie:1992qr, Kalinkin:2003uw, Ebert:2017udh, Ebert:2018dzs, Klimenko:2012qi, Klimenko:2012tk, Hands:2001aq, Hands:2005vn} but also of theoretical aspects of quantum field theory (e.g., renormalizability in the $1/N$ expansion or in a perturbative approach) \cite{Lenz:2020bxk,Hands:1991py, Stoll:2021ori}.
	In the mean-field approximation, the $1+1$-dimensional GN model exhibits a symmetric phase, a homogeneous symmetry-broken phase and an inhomogeneous phase, where the chiral condensate is an oscillating function of the spatial coordinate \cite{Thies:2003kk, Schnetz:2004vr, Thies:2006ti}.

	Recently, two independent mean-field studies of the $\mu$-$T$ phase diagram ($\mu$ denotes the baryon chemical potential) of the $2+1$-dimensional GN model were published \cite{Winstel:2019zfn, Narayanan:2020uqt, Buballa:2020nsi}.
	An important result of these works is that the existence of an inhomogeneous phase depends on the details of the regularization.
	When the regulator (lattice spacing or Pauli-Villars cutoff) is removed, the inhomogeneous phase disappears.
	
	In this work, we continue our previous lattice field theory investigations \cite{Winstel:2019zfn, Buballa:2020nsi} by studying the phase diagram of the $2+1$-dimensional GN model with an additional chiral chemical potential $\mu_{45}$.
\section{The $2+1$-dimensional GN model}
	The action of the $2+1$-dimensional GN model with baryon chemical potential $\mu$ as well as chiral chemical potential $\muff$ is given by
	\begin{align}
		\label{eq:fermi_action}
		S[\bar{\psi},\psi] = \int d^3x \, \bigg(\sum_{n=1}^{\Nf} \bar\psi_n \Big(\gamma_\nu \partial_\nu + \gamma_0 \mu + \gamma_{45} \gamma_0 \muff\Big) \psi_n - \cfrac{g^2}{2} \bigg(\sum_{n=1}^{\Nf} \bar\psi_n \psi_n\bigg)^2\bigg), 
	\end{align}
	where $\psi_n$ is a four-component massless fermion field, $\Nf$ is the number of flavors and $g^2$ is the coupling of the four-fermion interaction.
	A reducible representation of the Clifford algebra is chosen, where $\gamma_0$, $\gamma_1$ and $\gamma_2$ are block-diagonal. The upper two components of $\psi_n$ can be interpreted as being ``left-handed'', the lower two components as ``right-handed''.
	There are two linearly independent matrices $\gamma_4$ and $\gamma_5$, which anti-commute with $\gamma_0$, $\gamma_1$ and $\gamma_2$ and, thus, both fulfill the necessary properties of a suitable $\gamma_5$ matrix \cite{Scherer:2012nn, Gies:2009da}.
	Continuous symmetry transformations correspond to the generators $\mathds{I}_4$ and $\gamma_{45}$ \footnote{For $\Nf$ fermion fields one needs to consider tensor products with the generators of flavor rotations to get the full set of symmetry transformations.}, while discrete chiral symmetry transformations are given by either $\gamma_{4}$ or $\gamma_{5}$ \cite{Braun:2010tt,Gies:2010st}. 
	Since $\gamma_{45} = \diag(+\mathds{I}_2,\, -\mathds{I}_2)$, the chemical potential $\muff$ generates a chiral imbalance.
	More details can be found in \Rcite{Buballa:2020nsi} and will also be discussed in an upcoming publication.
	
	By introducing an auxiliary boson field $\sigma$ and integrating over the fermion fields, one obtains the effective action
	\begin{align}
		\label{eq:S_eff} \seff[\sigma] = \Nf \bigg(\frac{1}{2 \lambda} \int d^3x \, \sigma^2 - \ln\Big(\det(Q)\Big)\bigg) \quad , \quad Z = \int D\sigma \, e^{-\seff[\sigma]} ,
	\end{align}
	where
	\begin{equation}
		Q = \gamma_\nu \partial_\nu + \gamma_0 \mu + \gamma_{45} \gamma_0 \muff + \sigma
	\end{equation}
	is the Dirac operator and $\lambda = \Nf g^2$.
	The effective action is real-valued, if the boson field $\sigma$ is restricted to be constant in temporal direction (see appendix A of \Rcite{Buballa:2020nsi} for a proof).
	As $\seff \propto \Nf$, the limit $\Nf \rightarrow \infty$ reduces the relevant configurations in the path integral to the global minima of $\seff$, i.e., the mean-field approximation becomes exact in that limit.
	A Ward-Takahashi identity causes $\sigma$ to be proportional to the chiral condensate $\langle \bar{\psi}_n \psi_n \rangle$.
	Moreover, the effective action \cref{eq:S_eff} is invariant under $\sigma \rightarrow -\sigma$.
	Thus, $\sigma$ is an order parameter for the discrete chiral symmetry.
	
	The Dirac operator can be expressed in terms of irreducible representations of the Clifford algebra \cite{Urlichs:2007zz},
	\begin{equation}
		Q [\mu, \muff, \sigma] = \diag\left(Q^{(2)}[\mu + \muff,\sigma],\, \tilde{Q}{}^{(2)}[\mu - \muff,\sigma]\right), \,
		\label{eq:block_Q4}
	\end{equation}
	where $Q^{2}[\mu, \sigma]$ and $\tilde{Q}^{(2)}[\mu, \sigma]$ are defined as
	\begin{align}
		Q^{(2)}[\mu, \sigma] &= +\tau_2 (\partial_0 + \mu) + \tau_3 \partial_1 + \tau_1 \partial_2 + \sigma ,
		\\
		\tilde{Q}^{(2)}[\mu, \sigma] &= -\tau_2 (\partial_0 + \mu) - \tau_3 \partial_1 - \tau_1 \partial_2 + \sigma .
	\end{align}
	The Dirac operators $Q^{(2)}$ and $\tilde{Q}^{(2)}$ contain two inequivalent $2 \times 2$ representations, $\left(\gamma_0, \gamma_1, \gamma_2\right) = \left(\tau_2, \tau_3, \tau_1\right)$ and $\left(\gamma_0, \gamma_1, \gamma_2\right) = -\left(\tau_2, \tau_3, \tau_1\right)$, with $\tau_j$ denoting the Pauli matrices. They act on the ``left-handed'' and ``right-handed'' components introduced above.
	For $\muff = 0$ their chemical potentials are the same, but for $\muff \neq 0$ they differ, $\mu + \muff$ and $\mu - \muff$, indicating chiral imbalance.
	The determinants of $Q^{(2)}$ and $\tilde{Q}^{(2)}$ are invariant under $\sigma \rightarrow -\sigma$ as well as under $\mu \rightarrow -\mu$.
	A consequence of the latter is that $\det Q$ as well as effective action (\ref{eq:S_eff}) are invariant under the exchange of $\mu$ and $\muff$.
	This property is used to check our numerical results presented in section~\ref{SEC003}.
	
	One can show that 
	\begin{equation}
		\det Q^{(4)}[\mu, \sigma] = \left(\det Q^{(2)}[\mu, \sigma]\right)^2 = \left(\det \tilde{Q}^{(2)}[\mu, \sigma]\right)^2
	\end{equation}
	with $Q^{(4)}[\mu, \sigma] = Q[\mu, 0, \sigma]$ (see \Rcite{Buballa:2020nsi} for the proof). In consequence, the same phase diagram would be obtained if a Dirac operator
	\begin{equation}
		Q' [\mu, \muff, \sigma] = \diag\left(Q^{(4)}[\mu + \muff,\sigma],\, Q^{(4)}[\mu - \muff,\sigma]\right)
	\end{equation}
	would be used in the action. Then, the spinor fields would obtain an additional degree of freedom, independent of chirality, that one can interpret as ``isospin'' and $\muff$ would serve as an isospin chemical potential.
	We will explore this concept further in an upcoming publication.

\section{Lattice discretizations}
	We use lattice discretizations of \cref{eq:S_eff} similar to that discussed in section~4 of \Rcite{Buballa:2020nsi}. The main difference is that we use a lattice also in temporal direction instead of a superposition of plane waves.
	The cubic space-time volume is $\beta \times V=\beta \times L \times L$ and the corresponding lattice has spacing $a$ and $\Nt \times \Ns \times \Ns$ sites, i.e., $\beta = a\Nt$ and $L = a\Ns$.
	The boundary conditions are antiperiodic in temporal direction and periodic in the two spatial directions.

	For numerical computations it is convenient to work in momentum space.
	When using naive fermions, a weight function $\tilde{W}_2$ with specific properties has to be included in the interaction term of the GN model, to ensure the correct continuum limit (see \Rcite{Cohen:1983nr,Lenz:2020bxk, Buballa:2020nsi}, in particular appendix~2 of \Rcite{Lenz:2020bxk}).
	The Dirac operator is then
	\begin{align}
		Q(\xt{p},\xt{q}) = \ii \kron{\xt{p}}{\xt{q}} \sum_{\nu=0}^{2} \gamma_\nu \sin \big(p_\nu -  \kron{\nu}{0}\,\ii( \mu + \gamma_{45} \muff) \big) 
		+ \kron{p_0}{q_0} \tilde{W}_2(\x{p}-\x{q}) \tilde{\sigma}(\x{p}-\x{q}) ,
		\label{eq:diracop_disc}
	\end{align}
	where all quantities are expressed in units of $a$, i.e., dimensionless.
	In this work as well as in our previous work \cite{Buballa:2020nsi} we use and compare two weight functions corresponding to two different lattice discretizations,
	\begin{align}
		\tilde{W}'_2(\x{p}) = \prod_{\nu=1,2}  \tilde{W}'_1(p_\nu)\quad,\quad \tilde{W}'_1(p_\nu) = \frac{\cos(p_\nu) + 1}{2}
		\label{eq:W'_p}
	\end{align}
	and
	\begin{align}
		\tilde{W}''_2(\x{p}) = \prod_{\nu=1,2}  \tilde{W}''_1(p_\nu)\quad,\quad \tilde{W}''_1(p_\nu) = \Theta (\pi/2 - |p_\nu|),
		\label{eq:W''_p}
	\end{align}
	where $\Theta(p)$ is the Heaviside function.
\section{\label{SEC003}Numerical results}
	The phase diagram at $\muff = 0$ is known from \Rcite{Narayanan:2020uqt,Buballa:2020nsi}.
	There is a symmetric phase ($\sigma = 0$) at large $\mu$ and/or $T$ and a homogeneous symmetry-broken phase with a spatially constant field $\sigma = \bar{\sigma} \neq 0$ at small $\mu$ and small $T$.
	For $W_2 = W''_2$ and finite lattice spacing there is also an inhomogeneous phase, where $\sigma$ is a varying function of the spatial coordinates ($\sigma = \sigma(\x{x})$), at small $T$ between the symmetric and the homogeneous symmetry-broken phase.
	This phase, however, vanishes in the continuum limit.
	There is no inhomogeneous phase for $W_2 = W'_2$, neither at finite lattice spacing nor in the continuum limit.

	In this section we discuss numerical results for the two lattice discretizations (\ref{eq:W'_p}) and (\ref{eq:W''_p}) with the aim to extend the phase diagram to $\muff \neq 0$.
	Technical aspects like scale setting and tuning of the coupling constant are explained in \Rcite{Buballa:2020nsi}.
	As usual, all dimensionful quantities are expressed in units of $\sigma_0 = \sigma|_{\mu = 0, \muff = 0, T \approx 0}$, the vacuum expectation value of $\sigma$.
\subsection{The homogeneous phase diagram}
	\begin{figure}
		\centering
		\includegraphics[width=.7\textwidth]{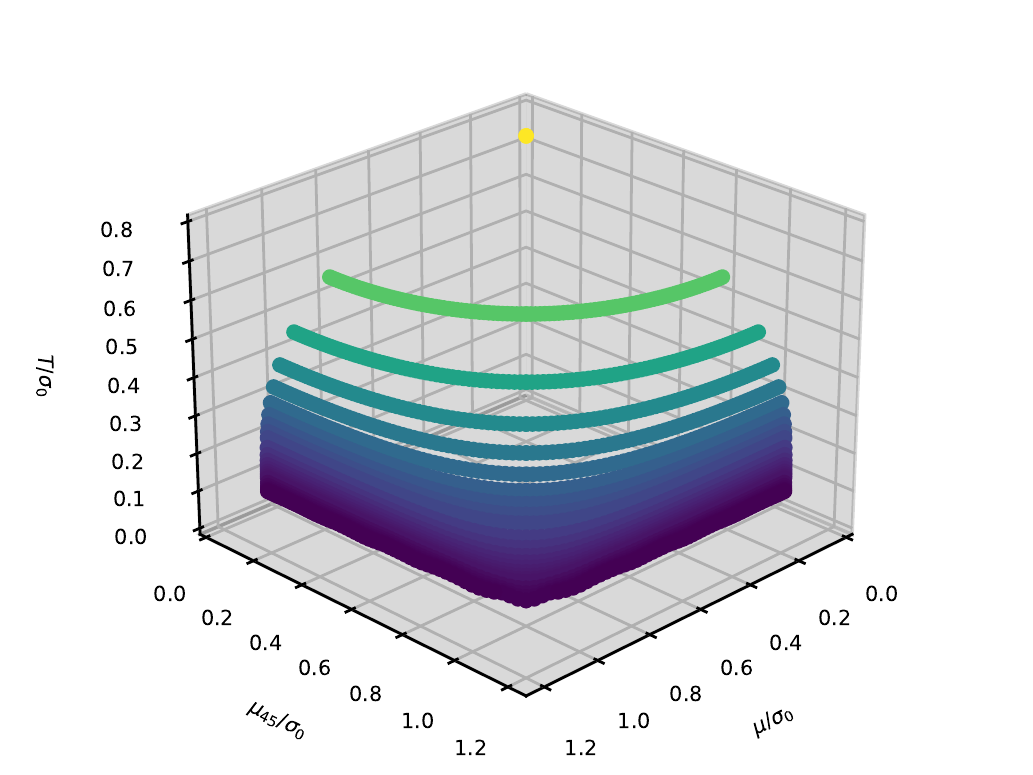}		
		\caption{\label{fig:iso2+1hom_pd}Phase diagram of the chirally imbalanced $2+1$-dimensional GN model with restriction to a homogeneous condensate in 3-dimensional $\mu$-$\muff$-$T$ space for $a \sigma_0 = 0.2327$ and $L \sigma_0 = 27.92$.}
	\end{figure}
	When only allowing a homogeneous condensate, i.e., $\tilde{\sigma}(\x{p}) = \bar{\sigma} \kron{\x{p}}{0}$, the lattice discretizations with $W'_2$ and $W''_2$ are identical (cf.\ \cref{eq:W'_p} and \cref{eq:W''_p} for $\x{p}=0$).
	The effective action at given $\mu$, $\muff$ and $T$ is a function of just one variable and can be minimized numerically in a straightforward way to determine the physically preferred value of $\bar{\sigma}$.

	\cref{fig:iso2+1hom_pd} shows the phase diagram in 3-dimensional $\mu$-$\muff$-$T$ space at finite lattice spacing and space-time volume. We note that these results reflect the previously discussed $\mu \leftrightarrow \muff$ symmetry of the effective action.
	The observed phase boundary for $\muff = 0$ is close to the continuum result from \Rcite{Klimenko:1987gi}, where small deviations are due to discretization errors and finite volume corrections. There are also slight differences to our lattice results from \Rcite{Buballa:2020nsi}, because of different discretizations in temporal direction.

	For large temperature, $T/\sigma_0 \gtrsim 0.4$, the phase boundary exhibits an approximate rotational symmetry in the $\mu$-$\muff$ plane, i.e., the phase boundary is crudely described by $\mu^2 + \muff^2 = \text{const}$.
	The behavior is completely different at low temperatures, $T/\sigma_0 \lesssim 0.1$, where the phase boundary resembles a square in the $\mu$-$\muff$ plane.
\subsection{Instabilities with respect to spatially inhomogeneous perturbations}
	\begin{figure}
		\begin{subfigure}[b]{0.5\textwidth}
			\includegraphics[width=\textwidth]{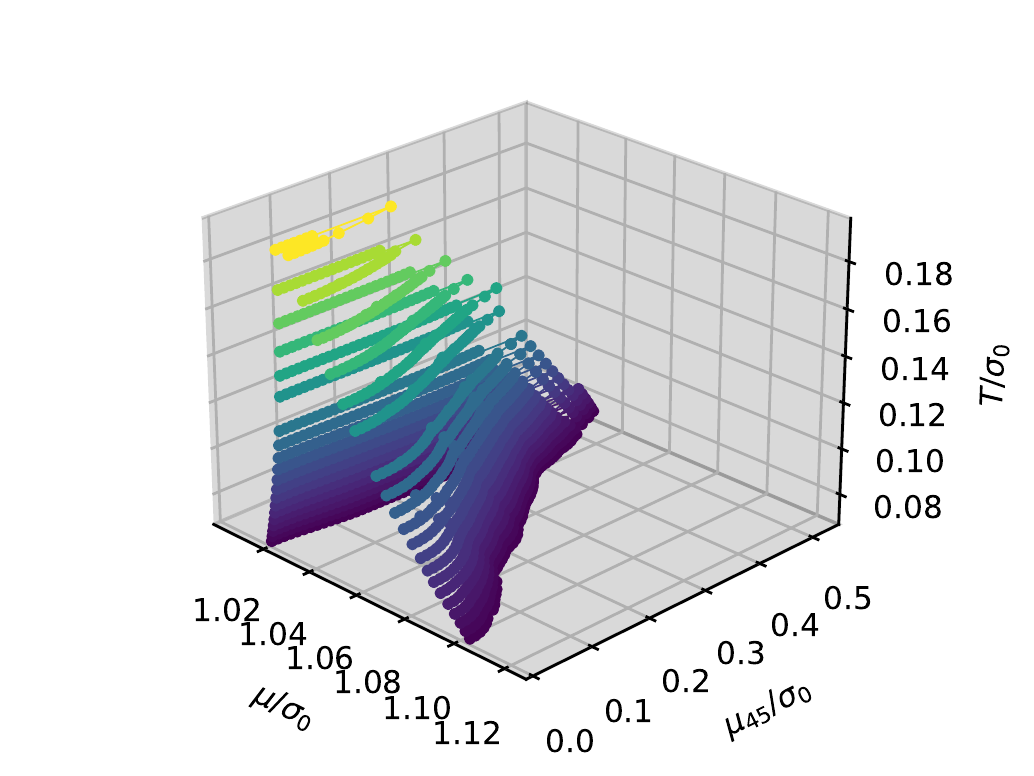}	
		\end{subfigure}
		\hfill
		\begin{subfigure}[b]{0.5\textwidth}
			\includegraphics[width=\textwidth]{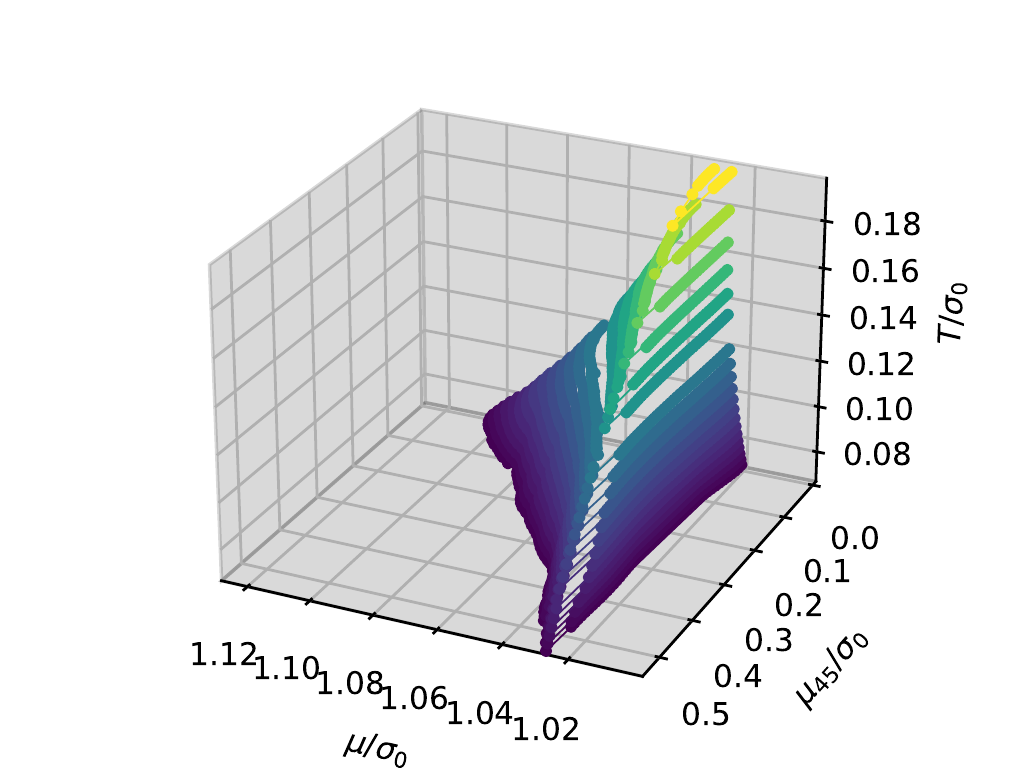}	
		\end{subfigure}
		
		\caption{\label{fig:iso_stab_analysis_3d}``Phase diagram'' of the chirally imbalanced $2+1$-dimensional GN model obtained via stability analyses in 3-dimensional $\mu$-$\muff$-$T$ space for $a \sigma_0 = 0.2327$ and $L \sigma_0 = 23.27$ and the discretization with $W_2 = W''_2$. The two plots show the same data from different angles.}
	\end{figure}
	To study possibly existing inhomogeneous phases, one has to allow for arbitrary modulations, i.e., non-vanishing $\sigma(\x{p})$, and to find the global minima of the effective action.
	In principle, such a minimization is possible within our lattice field theory approach, but it is a computationally extremely challenging task.
	For 1-dimensional modulations and $\muff = 0$ we were successfully using a local minimization algorithm with several randomly chosen initial field configurations \cite{Buballa:2020nsi}.
	In the near future we plan to use similar techniques to search for 2-dimensional modulations that are global minima, both at $\muff = 0$ and $\muff \neq 0$.

	In this work, a simpler strategy is applied and the existence of inhomogeneous phases is studied via stability analyses.
	For given $\mu$, $\muff$ and $T$ we check the stability of the homogeneous condensate $\bar\sigma$, as determined in the previous subsection, with respect to spatially inhomogeneous perturbations $\delta \tilde{\sigma}(\x{p})$ (for details we refer to \Rcite{Buballa:2020nsi}). If there is a perturbation $\delta \tilde{\sigma}(\x{p})$, which lowers the effective action compared to the homogeneous condensate, an inhomogeneous phase is indicated.
	In this way regions in the phase diagram can be mapped, which correspond to or are part of an inhomogeneous phase.

	Since there is no inhomogeneous phase in the continuum limit for chirally balanced matter \cite{Narayanan:2020uqt, Buballa:2020nsi}, we now explore, whether $\muff \neq 0$ might favor the existence of such a phase.
	We start with the discretization with $W_2 = W''_2$ and show in \cref{fig:iso_stab_analysis_3d} the boundaries of the instability region in 3-dimensional $\mu$-$\muff$-$T$ space at finite lattice spacing and space-time volume.
	The instability region resembles a tetrahedron and is adjacent to the homogeneously broken phase depicted in \cref{fig:iso2+1hom_pd}.
	Increasing $|\muff|$ seems to disfavor an inhomogeneous phase, because the instability region shrinks.

	Computations at several values of the lattice spacing suggest that the instability region vanishes in the continuum limit also at $\muff \neq 0$.
	This is supported by computations with the discretization with $W_2 = W'_2$, where no instability region is observed, neither at $\muff = 0$ (as discussed in \Rcite{Buballa:2020nsi}) nor at $\muff \neq 0$.
	These results will be shown and discussed in more detail in an upcoming publication.
\section*{Acknowledgements}
	We acknowledge useful discussions with L.~Kurth, A.~K\"onigstein, A.~Sciarra, M.~Thies and A.~Wipf. We thank A.~K\"onigstein for useful comments on the manuscript and for bringing \Rcite{Gies:2009da} to our attention.
	
	We acknowledge support by the Deutsche Forschungsgemeinschaft (DFG, German Research Foundation) through the CRC-TR 211 ``Strong interaction matter under extreme conditions'' -- project number 315477589-TRR 211. M.~Wagner acknowledges support by the Heisenberg Programme of the Deutsche Forschungsgemeinschaft (DFG, German Research Foundation) -- project number 399217702.
	M.~Winstel acknowledges support by the GSI Forschungs- und Entwicklungsvereinbarungen (GSI F\&E).
	Calculations on the GOETHE-HLR and on the on the FUCHS-CSC high-performance computers of the Frankfurt University were conducted for this research.
	We would like to thank HPC-Hessen, funded by the State Ministry of Higher Education, Research and the Arts, for programming advice.
\end{document}